\documentclass[aps,prb,twocolumn,showpacs,showkeys,floadfix]{revtex4}

\usepackage{bm}% bold math
\usepackage{amsmath}%ams
\usepackage{t1enc}              % polish letters
\usepackage[cp1250]{inputenc}   % polish keyboard
\usepackage[english]{babel}

\usepackage{graphicx}

\begin{document}

% Use the \preprint command to place your local institutional report
% number in the upper righthand corner of the title page in preprint mode.
% Multiple \preprint commands are allowed.
% Use the 'preprintnumbers' class option to override journal defaults
% to display numbers if necessary
%\preprint{}

%Title of paper
\title{Two-dimensional GaAs/AlGaAs superlattice structures for solar cell applications: ultimate efficiency estimation}

% repeat the \author .. \affiliation  etc. as needed
% \email, \thanks, \homepage, \altaffiliation all apply to the current
% author. Explanatory text should go in the []'s, actual e-mail
% address or url should go in the {}'s for \email and \homepage.
% Please use the appropriate macro foreach each type of information

% \affiliation command applies to all authors since the last
% \affiliation command. The \affiliation command should follow the
% other information
% \affiliation can be followed by \email, \homepage, \thanks as well.
\author{Jarosław W. Kłos}
\email{klos@amu.edu.pl}
%\homepage[]{Your web page}
%\thanks{}
\affiliation{Surface Physics Division, Faculty of Physics, Adam
Mickiewicz University, ul. Umultowska 85, 61-614 Poznań, Poland}
\affiliation{Department of Science and Technology, Linköping University
 601 74, Norrköping, Sweden
}

\author{Maciej Krawczyk}
%\email[]{Your e-mail address}
%\homepage[]{Your web page}
%\thanks{}
%\altaffiliation{}
%\affiliation{ }
\affiliation{Surface Physics Division, Faculty of Physics, Adam
Mickiewicz University, ul. Umultowska 85, 61-614 Poznań, Poland}

%Collaboration name if desired (requires use of superscriptaddress
%option in \documentclass). \noaffiliation is required (may also be
%used with the \author command).
%\collaboration can be followed by \email, \homepage, \thanks as well.
%\collaboration{}

\begin{abstract}
We calculate the band structure of a two-dimensional GaAs/AlGaAs superlattice and estimate the ultimate efficiency of solar cells using this type of structure for solar energy conversion. The superlattice under consideration consists of gallium arsenide rods forming a square lattice and embedded in aluminum gallium arsenide. The ultimate efficiency is determined versus structural parameters including the filling fraction, the superlattice constant, the rod geometry and the concentration of Al in the matrix material. The calculated efficiency of the superlattice proves to exceed the efficiency of each component material in the monolithic state in a wide range of parameter values.
\end{abstract}

% insert suggested PACS numbers in braces on next line
%Electron states at surfaces and interfaces,
%Electronic structure of nanoscale materials: clusters, nanoparticles,nanotubes, and nanocrystal
%Quantum wires
% insert suggested keywords - APS authors don't need to do this

%\maketitle must follow title, authors, abstract, \pacs, and \keywords
\maketitle

% body of paper here - Use proper section commands
% References should be done using the \cite, \ref, and \label commands
\section{Introduction}
In monolithic semiconductors with a single bandgap the ultimate efficiency, only due to limited absorption and to thermalization processes, is around $42\%$ (for the black-body radiation at $T_{s} = 6000$ K\cite{SQ}). The absorption of photons is incomplete because photons of energy below the bandgap width are not absorbed by the system. On the other hand, thermalization causes the surplus energy of electrons above the bottom of the conduction band or holes below the top of the valence band to dissipate in thermal contact with the crystal lattice. In ultimate efficiency calculations the utilized power is determined by the flux of absorbed photons and the width of the energy gap.

The efficiency taking into account recombination processes and the effect of thermodynamic losses is referred to as detailed balance efficiency. The recombination rate mainly depends on the temperature of the system. At ambient temperature only a minor part of excited carriers undergo radiative recombination to the ground state. Thermodynamic losses are a consequence of the fact that the chemical potential difference (for thermalized conduction band electrons and valence band holes) is lesser than the gap width\cite{wurfel}. Thus, the external voltage, even without load, is lower than that resulting from the gap width. With progressively increased load the current grows to saturate quickly, with a concurrent reduction of voltage at cell contacts. In the determination of detailed balance efficiency the utilized power is defined as the maximum power output at the optimum load. The theoretical limit of detailed balance efficiency is around $31\%$ (for the black-body radiation at $T_{s} = 6000$ K)\cite{SQ}.

The above-mentioned limitation of efficiency is referred to as the Shockley-Queisser limit. Many attempts have been made recently to find mechanisms and systems that would allow to overcome this constraint. Devices in which the Shockley-Queisser limit is outperformed are referred to as third-generation photovoltaic systems\cite{3g1,3g2,3g3}.

Photon absorption can be increased by using tandem cells - cascades of solar cells of successively narrowing bandgaps\cite{tandem,tandem2}. A similar effect can be obtained with solar-energy converters based on semiconductor superlattices\cite{nozik,green,kraw} with multiple bandgaps\cite{luq,34band,multiband,multiband2,multiband3}. Multiple-bandgap systems also allow to reduce thermalization losses.

Another method of increasing the efficiency is based on increasing the photocurrent or the photovoltage. The photocurrent can be increased by generation of extra carriers\cite{carrmul,carrmul2,carrmul3} in the inverse Auger effect\cite{invAug}. Otherwise lost in thermalization, the energy of highly excited carriers is thus utilized. The effectiveness of this process depends on carrier localization and is higher in dot and rod superlattices\cite{local}. The photovoltage can be increased in systems operating with hot electrons\cite{hc1,hc2,hc3}, in which energetically selective electrodes can capture highly excited carriers before they attain the thermodynamic equilibrium with the lattice through thermalization.

In this paper we determine the band structure of a two-dimensional GaAs/AlGaAs superlattice in the effective mass approximation\cite{bastard}.
Both GaAs and AlAs are well-known materials used widely in the fabrication of semiconductor heterostructures. Both compounds crystallize in the zinc blende structure and are characterized by similar lattice constants. In our calculations we use the envelope function approximation, which is widely applied for the description of conduction band bottom and valence band top for zinc blende crystallographic structures. We also omit the effect of stress on the matrix(AlGaAs)/rod(GaAs) interface. AlGaAs heterostructures have a relatively wide energy gap (1.4 eV - 2.1 eV) which makes them potentially suitable for intermediate band solar cells, where a wider gap (e.g. in comparison to the bulk Si) is required for achievement of high efficiency of solar radiation conversion.

In a wide range of structural parameter values minibands in the conduction band as well as those in the valence band overlap to a large extent. Only the lowest conduction miniband, narrow and quite strongly localized in potential wells formed by GaAs rods, is distinctly set apart\cite{kraw}. Higher minibands, which tend to lie above the matrix potential, are wide and overlap to form a continuous block. Even though some isolated minibands can be distinguished in valence band wells, the spacing between them is very narrow. Therefore, in our estimation of ultimate efficiency the band structure of the solar-energy converter is assumed to consist of a continuous valence band and a continuous conduction band with a single intermediate band formed by the lowest conduction miniband.

We determine the ultimate efficiency of a solar-energy converter of band structure as defined above, with the effects of incomplete photon absorption and carrier thermalization taken into account. The ultimate efficiency is examined versus structural parameters of the superlattice; these include the filling fraction, the superlattice constant, the rod geometry and the effective potential in the matrix material.

The paper is organized as follows. in the next section we discuss the geometry of the GaAs/AlGaAs superlattice under consideration and present the assumptions made in the effective mass approximation for conduction band electrons and for interacting heavy and light valence band holes. Then we present the plane-wave method used for the determination of the band structure of the superlattice. The following section discusses the assumptions made in the ultimate efficiency calculations and presents the method used for estimating the ultimate efficiency of a system with an intermediate band in the energy gap. Our numerical calculations are presented and discussed in a separate section. The paper is summed up in the Conclusions and supplemented by an Appendix that presents Fourier coefficients of those material parameters which are periodic functions of the position vector (the effective mass and the effective potential) for different rod geometries.

\section{Superlattice geometry and band structure}
\begin{figure}
\centering
\includegraphics[width=3in]{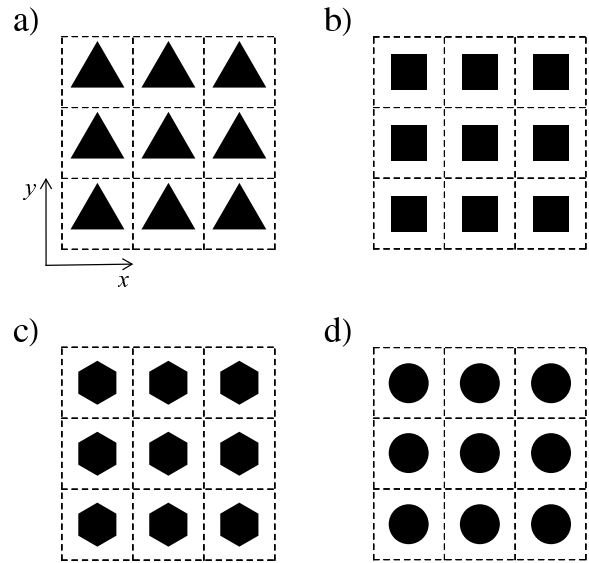}
\caption{\label{fig:f1}Structure of two-dimensional semiconductor superlattice, cross section perpendicular to rod axes (a nine-cell fragment of the infinite superlattice is shown). Black and white regions represent rod (well) and matrix (barrier) materials, respectively. Unit cells are delimited by dashed lines. Centers of rods of each geometry coincide with centers of unit cells. In each depicted superlattice the filling fraction value is $f = 0.25$. Rods of cross section in the form of (a) an equilateral triangle, (b) a square and (c) a regular hexagon are oriented so that symmetry axes of their cross sections coincide with symmetry axes of the unit cell.}
\end{figure}

Let us consider a two-dimensional superlattice formed by a system of GaAs rods disposed in sites of a square lattice and embedded in an AlGaAs matrix (see Fig.~\ref{fig:f1}). The rod axes are oriented along the $(001)$ direction (the $z$ direction) of the AlGaAs crystal lattice. The rods represent potential wells for both electrons and holes propagating in the $(x,y)$ plane perpendicular to the rod axes. The depth of the wells is determined by the concentration of Al in the matrix material. Increasing Al concentration causes the effective potential felt by conduction band electrons or valence band holes to rise or decrease, respectively, with consequent deepening of the wells.

Two-dimensional lattices formed by rods of different geometry are schematically represented in Fig.~\ref{fig:f1}. In this paper we consider rods of cross section in the form of an equilateral triangle (a), a square (b), a regular hexagon (c) or a circle (d). Rods are situated centrally in square unit cells of the superlattice and oriented in a manner which conserves as much as possible the symmetry of the lattice in the $x$ and $y$ directions. The filling fraction $f$, or the area ratio of the rod cross section to the unit cell, is a measure of the rod size. The maximum filling fraction value corresponds to the situation in which the rods reach the unit cell borders. For rods of triangular, hexagonal and circular cross-sectional geometry these maximum filling fraction values are $3\sqrt{3}/16$, $3\sqrt{3}/8$ and $\pi/4$, respectively.

The structure of energy minibands in the conduction band and in the valence band is determined here in the effective mass approximation, justified by the occurrence of a direct gap at point $\Gamma$ of the atomic band structure for low concentrations of Al in both the rods and the matrix. To meet the conditions of applicability of the effective mass approximation the Al concentration in the matrix material is only allowed to range from $0$ to $0.35$.

The following assumptions are made regarding electrons and holes:
\begin{itemize}
\item Interactions between the conduction band and the valence band are negligible, allowing independent determination of the miniband structure for electrons and holes.
\item Only the interactions between heavy and light hole bands in the valence band are taken into account.
\item The $z$ component of the wave vector is zero (electrons and holes propagate in the periodicity plane, $(x,y)$), which implies a spin degeneracy of heavy and light hole bands.
\end{itemize}
Electronic states are described by the Ben Daniel Duke equation\cite{duke}:
\begin{eqnarray}
      \left[-\alpha\left(\frac{\partial}{\partial x}\frac{1}{m^{*}({\bm r})}\frac{\partial}{\partial x}+
\frac{\partial}{\partial y}\frac{1}{m^{*}({\bm r})}\frac{\partial}{\partial y}\right)\right.\nonumber\\\Big.+E_{C}({\bm r})\Big]\Psi_{e}({\bm r})=E\Psi_{e}({\bm r}).   \label{eq:r1}
\end{eqnarray}
with effective mass $m^{*}({\bf r})$ (isotropic in a homogeneous medium) and effective potential $E_{C}({\bm r})$ determining the position of the conduction band bottom; $\Psi_{e}$ denotes the envelope wave function of a conduction band electron. The constant $\alpha =10^{-20}\hbar^{2}/(2 m_{e} e)\approx 3.80998$ ($m_{e}$ and $e$ denote the free electron mass and charge, respectively) allows to express the energy $E$ in electronvolts (eV) and the coordinates $x$, $y$, $z$ in angstroms ($\mathring{A}$). These material parameters vary in space with the periodicity of the superlattice:
\begin{eqnarray}
m^{*}({\bf r}+{\bf R})&=&m^{*}({\bf r}),\nonumber\\
E_{C}({\bf r}+{\bf R})&=&E_{C}({\bf r}),\label{eq:r2}
\end{eqnarray}
where ${\bf R}$ is a superlattice vector and ${\bf r}=(x,y)$ is the position vector.

The Schr\"{o}dinger equation of the envelope function of light and heavy hole states near the top of the valence band reads\cite{datta}:
\begin{equation}
\left( \begin{array}{cc}
\hat{P}+\hat{Q}&\hat{R}^{*}\\
\hat{R}&\hat{P}-\hat{Q}
\end{array} \right)
\left( \begin{array}{c}
\Psi_{lh}({\bm r})\\
\Psi_{hh}({\bm r})
\end{array}
\right)
=E\left( \begin{array}{c}
\Psi_{lh}({\bm r})\\
\Psi_{hh}({\bm r})
\end{array}
\right),\label{eq:r3}
\end{equation}
where $\Psi_{lh}({\bm r})$ and $\Psi_{hh}({\bm r})$ are envelope functions for light and heavy holes, respectively, and operators $\hat{P}$, $\hat{Q}$ and $\hat{R}$ have the form:
\begin{eqnarray}
P&=&E_{V}({\bm r})+\alpha \left(
\frac{\partial}{\partial x}\gamma_{1}({\bm r})\frac{\partial}{\partial x}+
\frac{\partial}{\partial y}\gamma_{1}({\bm r})\frac{\partial}{\partial y}
\right),
\nonumber\\
Q&=&\alpha \left(
\frac{\partial}{\partial x}\gamma_{2}({\bm r})\frac{\partial}{\partial x}+
\frac{\partial}{\partial y}\gamma_{2}({\bm r})\frac{\partial}{\partial y}
\right),
\nonumber\\
R&=&\alpha\sqrt{3}\left[
- \left(
\frac{\partial}{\partial x}\gamma_{2}({\bm r})\frac{\partial}{\partial x}-
\frac{\partial}{\partial y}\gamma_{2}({\bm r})\frac{\partial}{\partial y}
\right)\right.\nonumber\\
& & +\left. i \left(
\frac{\partial}{\partial x}\gamma_{3}({\bm r})\frac{\partial}{\partial y}+
\frac{\partial}{\partial y}\gamma_{3}({\bm r})\frac{\partial}{\partial x}
\right)
\right].\label{eq:r4}
\end{eqnarray}

Luttinger parameters $\gamma_{1}$, $\gamma_{2}$, $\gamma_{3}$, describing, respectively, the effective masses $1/(\gamma_{1}+\gamma_{2})$ and $1/(\gamma_{1}-\gamma_{2})$ of light and heavy holes near the point $\Gamma$ of the atomic lattice, are, like the position of the valence band top $E_{V}$, periodic in the superlattice structure.
\begin{eqnarray}
\gamma_{\beta}({\bf r}+{\bf R})&=&\gamma_{\beta}({\bf r}),\nonumber\\
E_{V}({\bf r}+{\bf R})&=&E_{V}({\bf r}),\label{eq:r5}
\end{eqnarray}
where label $\beta$ denotes: $1,2,3$.

The effective mass of carriers and the effective potential in which they move depend on the space-variable chemical composition of AlGaAs (alloy between GaAs and AlAs). Rods, which represent potential wells, are made of GaAs; the matrix material, representing a barrier for electrons and holes, is AlGaAs. The following empirical formulae, obtained by linear extrapolation of values of material parameters in GaAs and AlAs, allow to estimate their values in the matrix material\cite{param,param2}:
\begin{eqnarray}
E_{C}&=&0.944 d,\nonumber\\
E_{V}&=&1.519+0.75 d,\nonumber\\
m^{*}&=&0.067 + 0.083 d,\nonumber\\
\gamma_{1}&=&6.85- 3.40 d,\nonumber\\
\gamma_{2}&=&2.10- 1.42 d,\nonumber\\
\gamma_{3}&=&2.90- 1.61 d,\label{eq:r6}
\end{eqnarray}
where $d$ is the concentration of Al in gallium arsenide.

The expansion in the plane-wave basis of the envelope function for holes and electrons and the Fourier expansion of the material parameters can be formally written as:
\begin{eqnarray}
\Psi_{\beta_{1}}({\bm r})&=&\sum_{{\bm G}}\phi_{\beta_{1}}^{\bm G}e^{i ({\bm G}+{\bm k})\cdot{\bm r}},\nonumber\\
E_{\beta_{2}}({\bm r})&=&\sum_{{\bm G}}E_{\beta_{2}}^{\bm G}e^{i {\bm G}\cdot{\bm r}},\nonumber\\
w({\bm r})=1/m^{*}({\bm r})&=&\sum_{{\bm G}}w^{\bm G}e^{i {\bm G}\cdot{\bm r}},\nonumber\\
\gamma_{\beta_{3}}({\bm r})&=&\sum_{{\bm G}}\gamma_{\beta_{3}}^{\bm G}e^{i {\bm G}\cdot{\bm r}},\label{eq:r7}
\end{eqnarray}
where the labels $\beta_{1}$, $\beta_{2}$, $\beta_{3}$ refer to $e,hh,lh$; $C,V$; $1,2,3$, respectively; $\phi_{\beta_{1}}^{\bm G}$ have the sense of Fourier coefficients of the periodic factor of the envelope function (which has the same periodicity as the superlattice). Fourier coefficients of the material parameters can be found analytically from the formula:
\begin{eqnarray}
F^{\bm G}=\frac{1}{S}\int_{S}f({\bm r})e^{-i{\bm G}\cdot{\bm r}}d{\bm r},\label{eq:r8}
\end{eqnarray}
for each of the four rod geometries under consideration. Symbols $f({\bm r})$ and $F^{\bm G}$ denote, respectively, the periodic material parameter (effective mass or effective potential) and the corresponding Fourier coefficient for a plane wave of wave vector equal to vector ${\bm G}$ of the reciprocal superlattice. Parameter $S$ represents the area of a superlattice cell. For the explicit form of the Fourier coefficients (\ref{eq:r7}) for rods of triangular, square, hexagonal or circular cross section, see the Appendix.

By substitution of expansions (\ref{eq:r7}) and (\ref{eq:r8}) in the Schrodinger equations (\ref{eq:r1}) and (\ref{eq:r3}) for electrons and holes, respectively, we obtain systems of equations in the form of an eigenvalue problem, the solution of which yields Fourier coefficients of the periodic factor of the envelope function and the carrier energy. For electronic states we get:
\begin{eqnarray}
\sum_{{\bm G'}}\Big[\alpha\left({\bm G}+{\bm k}\right)\cdot\left({\bm G'}+{\bm k}\right)w^{{\bm G'}-{\bm G}}
\nonumber\\
+E_{C}^{{\bm G'}-{\bm G}}\Big]\phi_{e}^{{\bm G'}}=E\phi_{e}^{{\bm G}}\label{eq:r9}
\end{eqnarray}
In the case of interacting light and heavy holes the eigenvalue problem reads:
\begin{eqnarray}
\sum_{{\bm G'}}\Big[\left(A_{1}+A_{2}\right)\phi_{lh}^{{\bm G'}}+\left(B_{1}-iB_{2}\right)\phi_{hh}^{{\bm G'}}\Big]=E\phi_{lh}^{{\bm G}},\nonumber\\
\sum_{{\bm G'}}\Big[\left(B_{1}+iB_{2}\right)\phi_{lh}^{{\bm G'}}+\left(A_{1}-A_{2}\right)\phi_{hh}^{{\bm G'}}\Big]=E\phi_{hh}^{{\bm G}}\label{eq:r10}
\end{eqnarray}
where $A_{1}, A_{2}$ and $B_{1}, B_{2}$ are expressed as follows:
\begin{eqnarray}
A_{1}&=&-\alpha\gamma_{1}^{{\bm G}-{\bm G'}}\left({\bm G}+{\bm k}\right)\cdot\left({\bm G'}+{\bm k}\right)-E_{V}^{{\bm G}-{\bm G'}},\nonumber\\
A_{2}&=&-\alpha\gamma_{2}^{{\bm G}-{\bm G'}}\left({\bm G}+{\bm k}\right)\cdot\left({\bm G'}+{\bm k}\right),\nonumber\\
B_{1}&=&-\alpha\sqrt{3}\gamma_{2}^{{\bm G}-{\bm G'}}\left[(G_{x}+k_{x})(G'_{x}+k_{x})\right.\nonumber\\&&-\left.(G_{y}+k_{y})(G'_{y}+k_{y})\right],\nonumber\\
B_{2}&=&\alpha\sqrt{3}\gamma_{3}^{{\bm G}-{\bm G'}}\left[(G_{x}+k_{x})(G'_{y}+k_{y})\right.\nonumber\\&&+\left.(G'_{x}+k_{x})(G_{y}+k_{y})\right]\label{eq:r11}
\end{eqnarray}

\begin{figure}
\centering
\includegraphics[width=3in]{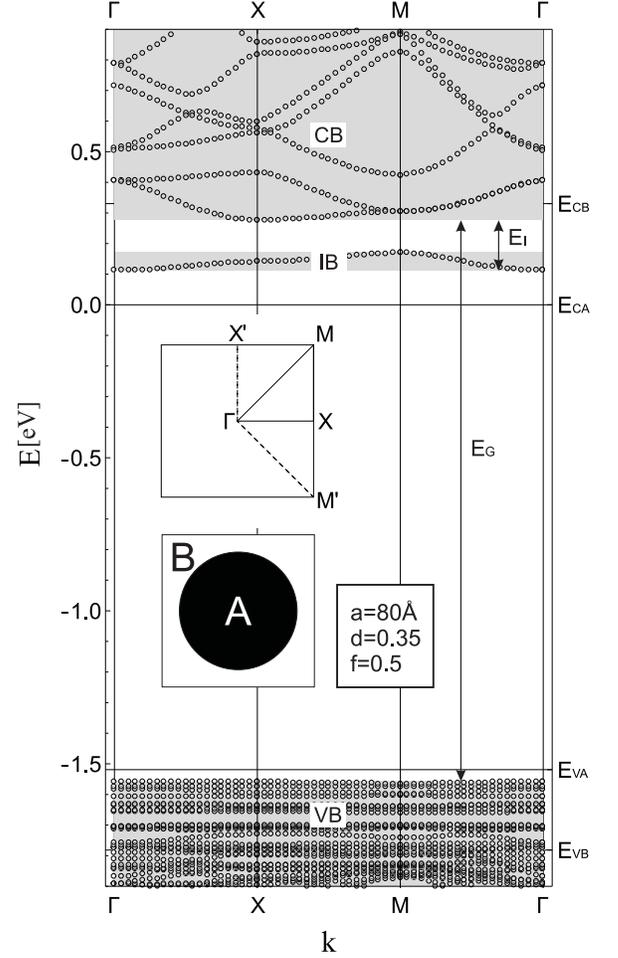}
\caption{\label{fig:f2}Energy spectrum of semiconductor superlattice formed by cylindrical rods of GaAs (A) embedded in a matrix of Al$_{d}$Ga$_{1-d}$As ($d = 0.35$) (B), with filling fraction $f = 0.5$ and superlattice constant $a = 80\mathring{A}$. The dispersion relation is plotted along the $\Gamma-X-M-\Gamma$ path shown in the inset. The horizontal lines indicate the position of the conduction band bottom in the rod and matrix materials ($E_{CA}$ and $E_{CB}$, respectively) and the position of the valence band top in the rod and matrix materials ($E_{VA}$ and $E_{VB}$, respectively). The reference energy level is $E_{CA} = 0$. The lowest miniband, detached from the conduction band, is regarded as an intermediate level that opens an extra channel for carrier transitions between the valence band and the conduction band. The ultimate efficiency of the solar cell is determined by the gap $E_{G}$ between the conduction band and the valence band, and the distance $E_{I}$ between the bottom of the intermediate miniband and the conduction band.}
\end{figure}

Figure~\ref{fig:f2} presents the band structure of a superlattice consisting of cylindrical rods (material A) embedded in a matrix (material B) with Al concentration $d = 0.35$, for a filling fraction $f = 0.5$ and a superlattice constant $a = 80 \mathring{A}$. The electron energy is referred to the energy $E_{CA} = 0$ of the conduction band bottom in the rod material; $E_{CB}$ is the position of the conduction band bottom in the matrix material; $E_{VA}$ and $E_{VB}$ denote the position of the valence band top in the rod and matrix materials, respectively. The dispersion relation for minibands is plotted along the high-symmetry path $\Gamma-X-M-\Gamma$ shown in the inset.

In the $(x,y)$ plane the mass of light and heavy holes is greater than that of electrons, which implies valence minibands narrower than conduction minibands. As the miniband structure is generated by both the light and heavy hole systems, valence minigaps are also narrower than conduction minigaps. Interactions between the systems of light and heavy holes can cause additional narrow minigaps to open and minibands to split as a consequence of the lifting of the degeneracy of the two systems. Numerous very narrow valence minigaps are seen to occur in the well ($E_{VB}<E<E_{VA}$), and valence minibands to overlap below the barrier height ($E<E_{VB}$) (Fig.~\ref{fig:f2}). To simplify the ultimate efficiency calculation we assume the valence band is uniform (i.e. the minibands are merged) in the superlattice structures considered, and the highest miniband marks the top of the valence band.

In a wide range of model parameters only the first (lowest) conduction miniband is distinctly detached from the band. This isolated miniband lies in the potential well ($E_{CA}<E<E_{CB}$). Minibands above the barrier height ($E>E_{CB}$) are inclined to overlap and tend to form a uniform block.

In a very rough approximation, the band structure of the superlattice can be regarded as composed of a valence band, or block, ($VB$) and a block of overlapping conduction minibands, with a wide gap $E_{G}$ between the two blocks. The lowest, detached conduction miniband can be regarded as an intermediate band ($IB$) within the gap. The energy $E_{I}$ indicated in Fig.~\ref{fig:f2} is the shift between the bottom of the $IB$ and the block of overlapping conduction minibands. The latter shall be henceforth referred to as the conduction band ($CB$).

\begin{figure}
\centering
\includegraphics[width=3in]{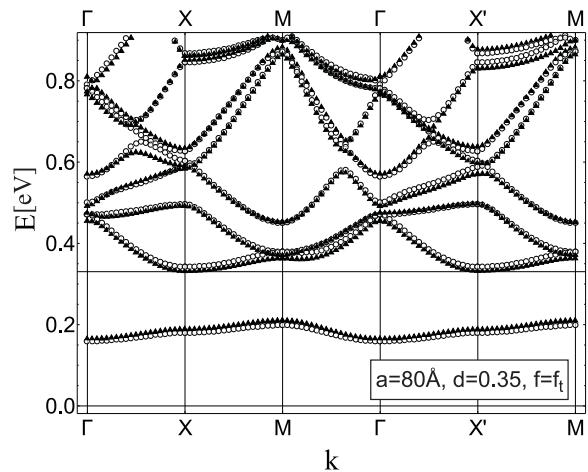}
\caption{\label{fig:f3}Energy spectrum of conduction minibands of superlattices with rods of circular or triangular cross section (circles and filled triangles, respectively). Plotted along the path $\Gamma-X-M-\Gamma-X'-M$ shown in Fig.~\ref{fig:f2}, the dispersion relation shows non-equivalence of Brillouin zone points $X$ and $X'$ in the superlattice with rods of triangular cross section. The following parameter values are assumed in both cases: superlattice constant $a = 80\mathring{A}$, matrix Al concentration $d = 0.35$ and filling fraction $f_{t} = 3\sqrt{3}/16$.}
\end{figure}

If the symmetry of the rod cross section is lower than that of a square, points $X$ and $X'$ in the Brillouin zone, as well as points $M$ and $M'$, are not equivalent (see Fig.~\ref{fig:f2}, inset). In this study rods of triangular and hexagonal cross section are oriented against the superlattice in the manner shown in Fig.~\ref{fig:f1}. Propagation of an electron wave along the $\Gamma-X$ direction is easily seen not to be equivalent to its propagation along $\Gamma-X'$. On the other hand, propagation along $\Gamma-M$ and $\Gamma-M'$ is identical. To examine the effect of the rod symmetry on the superlattice spectrum we have compared the dispersion relations for structures with rods of circular and triangular cross section. Figure~\ref{fig:f3} shows the dispersion relation along the path $\Gamma-X-M-\Gamma-X'-M$ in superlattices with rods of both geometries. The calculations were performed for a filling fraction $f = 3\sqrt{3}/16$, the value corresponding to the triangular rods touching the borders of the superlattice cell. In the superlattice with rods of triangular cross section the dispersion relation near points $X$ and $X'$ is only seen to differ visibly in higher minibands. Obviously, no such differences are found in the spectrum of the superlattice with cylindrical rods. However, the two spectra do not differ much. The only differences are the lowest conduction miniband (the $IB$) slightly shifted to higher energy and the $CB$ slightly shifted to lower energy in the superlattice with rods of triangular cross section.

\section{Estimation of ultimate efficiency of systems with IB}

In our estimation of ultimate efficiency $\eta$ we made the following assumptions regarding the radiation of the source and the solar cell. 
\begin{itemize}
\item The cell absorbs the amount of black-body radiation at a temperature corresponding to the maximum of the solar spectrum (the value assumed in the calculations is $T_{s} = 5760 K$). This assumption simplifies our calculation and could affect the results only if optical transitions in the solar cell
structure coincide with windows in the terrestrial sun spectrum.
\item The temperature of the solar cell is $T = 0 K$, which implies no recombination processes. The absence of recombination processes allows us to discard the impact of geometry of the solar cell and its alignment with respect to the incident solar radiation. 
\end{itemize}
The assumptions listed below describe the features of the intermediate band photoconverter which is able to reach the maximal utilization of photon energy in the absence of carrier multiplication processes. The computed ultimate efficiency results only from selective absorption in a given band structure of the photoconverter.
\begin{itemize}
\item Electronic transitions always occur through energy gaps of the greatest width available to the given photon energy. This implies minimized losses in the thermalization process.
\item If the photon energy is high enough for an electron to cross the gap, the probability of the transition is 100\%. The selection rules for optical transitions\cite{opttrans} are not taken into account.
\item The energy of excited carriers is not utilized until the electron crosses the gap to reach the conduction band as a result of cascade transitions via intermediate bands. Then each electron brings an initial energy equal to the width of the gap $E_{G}$ between the VB and the CB (Fig.~\ref{fig:f2}).
\item The flux of electrons excited to the conduction band via intermediate bands is equal to the lowest flux of photons absorbed in a single gap in the cascade: $VB-IB-CB$.
\item Each absorbed photon generates a single electron-hole pair.
\end{itemize}
A further increase of efficiency can by achieved by hot carrier extraction (when photon energy $E>E_{G}$) or by carrier multiplication (inverse Auger process). Hence, the assumptions presented above give the least restrictive limitation for efficiency in the absence of carrier multiplication and hot carrier utilization. The definition of ultimate efficiency we used results mainly from the photoconverter band structure and does not take into account the external geometry of the system (the light management and concentration), the mismatch to the value of load (the shift of the electrochemical potentials), etc.

The power of radiation per unit area of the solar cell is described by Planck's distribution for black-body radiation\cite{SQ}:
\begin{equation}
P_{in}=2\pi^{5}(k_{B}T_{s})^{4}/15h^{3}c^{2},\label{eq:r12}
\end{equation}
where $k_{B}$ is the Boltzmann constant, $h$ is the Planck constant and $c$ is the velocity of light.
Three ranges of photon energy E can be distinguished in the considered case with a single IB. These three ranges correspond to the following three types of carrier transitions (Fig.~\ref{fig:f2}):
\begin{itemize}
\item between the $IB$ and the $CB$: $E_{I}<E<E_{G}-E_{I}$,
\item from the $VB$ to the $IB$: $E_{G}-E_{I}<E<E_{G}$,
\item from the $VB$ directly to the $CB$: $E>E_{G}$.
\end{itemize}
The flux of photons absorbed by a unit area of the cell per unit time is governed by the following general formula for each of the above transitions:
\begin{eqnarray}
I(E_{1},E_{2})&=&2\pi(k_{B}T_{s})^{3}/h^{3}c^{2}\int_{\xi(E_{1})}^{\xi(E_{2})}\frac{\xi^{2}d\xi}{e^{\xi}-1},\nonumber\\
&&\xi(E)=E/k_{B}T_{s},\label{eq:r13}
\end{eqnarray}
where $E_{1}$ and $E_{2}$ denote the above-specified limits of the energy range for the given transition. The utilized power per unit area of the cell\cite{SQ}:
\begin{eqnarray}
\begin{array}{l}
P_{out}=E_{G}\Big[I(E_{G},\infty)\Big.\\+\Big.\min\left(I(E_{G}-E_{I},E_{G}),I(E_{I},E_{G}-E_{I})\right)\Big]
\end{array}\label{eq:r14}
\end{eqnarray}
has two components. One, $I(E_{G},\infty)$, is related to direct carrier transitions between the $VB$ and the $CB$. The other component is the minimum value of two streams, $I(E_{G}-E_{I},E_{G})$ and $I(E_{I},E_{G}-E_{I})$, from cascade-like transition via the $IB$.
The ultimate efficiency is defined as a ratio of utilized output power to incident power:
\begin{eqnarray}
\eta=\frac{P_{out}}{P_{in}}.\label{eq:r15}
\end{eqnarray}

\begin{figure}
\centering
\includegraphics[width=3in]{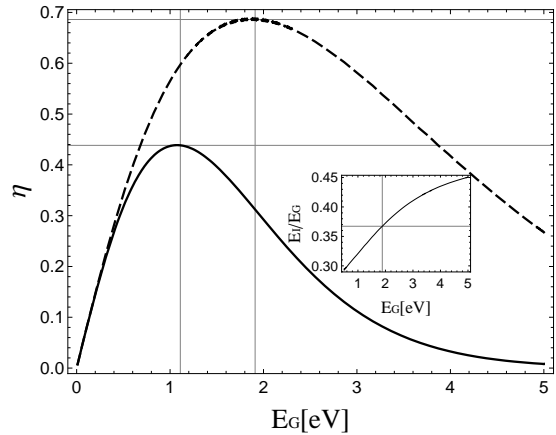}
\caption{\label{fig:f4}Ultimate efficiency versus bandgap width $E_{G}$ for two solar cells, without (solid line) and with (dashed line) an $IB$ shifted by $E_{I}$ from the conduction band edge. The efficiency of the system with $IB$ was determined for optimum position of the $IB$ in the bandgap (see the inset), corresponding to equal fluxes of photons absorbed in transitions from $VB$ to $IB$ and from $IB$ to $CB$. Indicated in the plot, the maximum efficiency values for the systems without and with $IB$ are $42\%$ (for $E_{G} = 1.1$ eV) and $68\%$ (for $E_{G} = 1.9$ eV), respectively.}
\end{figure}

Figure~\ref{fig:f4} presents the ultimate efficiency plotted versus the energy gap width $E_{G}$ for two solar cells, one without and the other with an $IB$ in the bandgap. The assumed position of the $IB$ corresponds to the situation in which the flux of photons generating transitions between the $VB$ and the $IB$ is equal to the flux of photons generating transitions between the $IB$ and the $CB$. This means both fluxes are fully used for carrier transitions between the $VB$ and the $CB$. The inset in Fig.~\ref{fig:f4} shows the optimum relative position of the $IB$ in the gap between the $VB$ and the $CB$ plotted versus the gap width $E_{G}$. The superimposed plot for the two systems under consideration indicates the ultimate efficiency is substantially higher in the system with $IB$, in which case the maximum ultimate efficiency, about $68\%$, corresponds to a gap width $E_{G} = 1.9$ eV (in the solar cell without IB the maximum efficiency is $42\%$ and corresponds to $E_{G} = 1.1$ eV).

These properties suggest AlGaAs could be a good candidate for applications in solar cells based on materials with an $IB$ in the bandgap, as the bandgap width in the monolithic materials ranges from 1.52 eV in GaAs to 3.21 eV in AlAs. However, in the superlattice model under consideration the $IB$, being the lowest conduction miniband, lies in the potential well between the conduction band bottom in the rod material (GaAs) and in the matrix material (AlGaAs). Consequently, the distance $E_{I}$ between the $IB$ and the block of overlapping conduction minibands is limited and far from the optimum value; this represents a major obstacle in attaining maximum ultimate efficiency. However, the results presented in the next section indicate a sensible increase in ultimate efficiency with respect to its values in the monolithic materials without $IB$ (AlGaAs or GaAs) in a wide range of superlattice geometry parameters.

\section{Results}
We have determined the ultimate efficiency for structures depicted in Fig.~\ref{fig:f1} and examined it versus the following structural parameters: the filling fraction $f$, the superlattice constant $a$ and the concentration $d$ of aluminum in the matrix material, the latter parameter determining the depth of the effective potential well. The ultimate efficiency has been plotted versus each of these parameters, with the others fixed. We realize that a comprehensive study should take into account the simultaneous interplay of all mentioned parameters but such investigation would be very difficult in analysis and require an enormous amount of computational time. The values of fixed parameters: $f=0.5$, $a = 80 \mathring{A}$ , $d=0.35$, were chosen arbitrarily. We tried, however, to give some convincing justification for such choice. The filling fraction $f=0.5$ describes the system in which the volume of the matrix and the rod materials are equal; for the superlattice constant
 $a \approx 80 \mathring{A}$ the band spectrum looks quite regular (the minibands and the minigaps are of comparable width); increasing the depth of the wells and the width of the IB (which results in higher efficiency), the range of Al concentration $d$ in AlGaAs is limited to $d=0.35$ for the envelope function approximation to be applicable.

\begin{figure}
\centering
\includegraphics[width=3in]{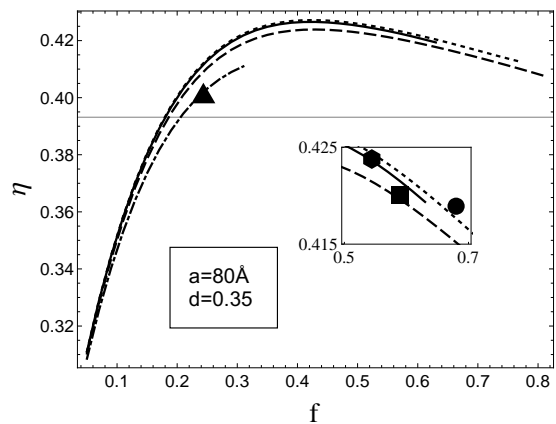}
\caption{\label{fig:f5}Ultimate efficiency $\eta$ versus filling fraction $f$, as calculated for superlattice constant $a = 80 \mathring{A}$ and matrix Al concentration d = 0.35. Each line refers to a different rod geometry, as indicated by labels of respective shape. The horizontal line indicates the efficiency of monolithic rod material, GaAs.}
\end{figure}

Figure~\ref{fig:f5} shows the ultimate efficiency $\eta$ versus the filling fraction $f$ for superlattices with rods of triangular, square, hexagonal or circular cross section. 
We fixed the superlattice constant at $a = 80 \mathring{A}$ and the matrix Al concentration at d = 0.35.
In the range of very low filling fraction values the effective potential wells formed by the rods become too narrow to bind electronic states. Pushed out of the well, the conduction miniband merges into the continuum of higher minibands; as a result, the superlattice band structure resembles that of monolithic AlGaAs with a continuous conduction band. The gap between the conduction band and the lowest miniband closes at a filling fraction value of $0.05$, approximately. Note the efficiency of the superlattice with a minimum width of the gap between the $IB$ and the $VB$ is relatively high, about $31\%$, against the $26.8\%$ ultimate efficiency of a homogeneous matrix material (AlGaAs). This rather substantial difference can be explained as follows. When the effective potential well is narrow the position of the $IB$ varies the most; the energy of the $IB$ increases relatively fast with decreasing filling fraction. However, as the $IB$ is pushed out of the well, it grows wider, and at the moment of closing of the gap that separates it from the $VB$ its width is quite significant and its bottom lies below the effective potential of the matrix. In this limiting case $E_{I}$ (cf. Fig.~\ref{fig:f2}) is substantially different from zero, while $E_{G}$ is close to the width of the gap in monolithic AlGaAs. This affects significantly the ultimate efficiency in the model considered.

The ultimate efficiency increases rapidly as the filling fraction continues to grow. At $f = 0.18$ the efficiency of a superlattice-based solar-energy converter is equal to that of a solar cell made of monolithic GaAs (the horizontal line in Fig.~\ref{fig:f5}). This steep increase in efficiency is due to the widening of the potential wells formed by the rods, and the concurrent narrowing of the barriers formed by the matrix material. As the wells grow wider, the first conduction miniband and the first valence miniband move towards the well bottom, which results in a reduced $E_{G}$. Consequently, transitions omitting the $IB$ become more intense, resulting in an increased efficiency. However, narrowing miniband spacing implies reduced distance between the $IB$ and the $CB$. This competitive effect reduces the rate of transitions via $IB$. The ultimate efficiency reaches maximum at filling fraction $0.42$, approximately. Further widening of the wells, or thickening of rods at the cost of the matrix material, leads to a decrease in efficiency.

Note the filling fraction cannot reach 1 in structures with rods of triangular, hexagonal or circular cross section. Its limit values, $3\sqrt{3}/8$ and $\pi/4$, for rods of hexagonal and circular cross section, respectively, correspond to the situation in which the rods touch at the borders of superlattice unit cells. In the case of rods of triangular cross section the limit value $f = 3\sqrt{3}/16$ assumed in the calculations corresponds to rods touching the unit cell borders rather than one another, and is motivated by technical reasons and otherwise very complex calculations.

The minibands grow in width as the filling fraction continues to increase. In the superlattice with square cross-section rods the gap between the $IB$ and the $CB$ closes at $f = 0.81$. The bottom of the continuous CB formed in this way is very close to the conduction band bottom in GaAs. However, the ultimate efficiency values of monolithic GaAs and a zero-gap superlattice differ by $1\%$, approximately. This quite significant difference is due to the fact that as long as the gap remains open the conduction band bottom refers to the block of higher minibands. It is from this level that the energy of excited electrons is utilized in our model. When the $IB$ merges into the $CB$ the energy of utilized carriers is reduced by the width of the $IB$.

As long as the filling fraction value is fixed, the geometry of the rods proves of little importance for the efficiency. The efficiency is seen to increase as the shape of the rods becomes more compact and regular: the superlattice with cylindrical rods has a slightly higher efficiency than the superlattices with rods of triangular or square cross section, the approximate maximum differences being $1.2\%$ and $0.4\%$, respectively.

\begin{figure}
\centering
\includegraphics[width=3in]{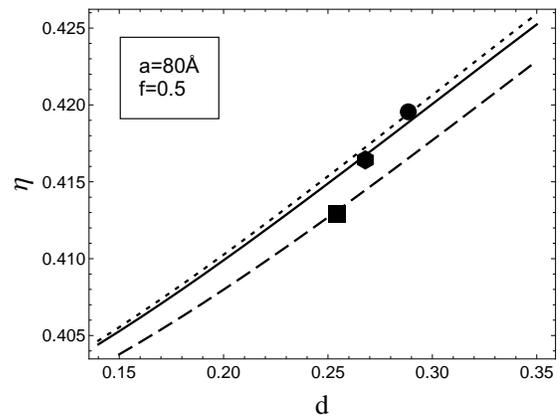}
\caption{\label{fig:f6}Ultimate efficiency $\eta$ versus Al concentration $d$ in the matrix material ($d$ determines the depth of potential wells formed by the rods with respect to the barriers formed by the matrix material). Calculations were performed for superlattice constant $a = 80\mathring{A}$ and filling fraction $f = 0.5$. Plots obtained for superlattices with rods of square, hexagonal and circular cross section are labeled with respective geometric figures.}
\end{figure}

Figure~\ref{fig:f6} shows the ultimate efficiency plotted versus the concentration $d$ of aluminum in the matrix material for superlattices with rods of square, hexagonal and circular cross section. The Al concentration in the matrix material determines the depth of the effective potential wells felt by carriers. The calculations were performed for the filling fraction fixed at $f = 0.5$ and superlattice constant $a = 80 \mathring{A}$. For $d$ values around 0.14 the wells are too shallow to effectively bind electrons or holes, and the spectrum of the superlattice resembles that of monolithic GaAs. The intermediate band in these conditions is relatively wide, its bottom very close to that of the conduction band in monolithic GaAs, and the absolute gap between $IB$ and the $VB$ extremely narrow. The efficiency exceeds by $1\%$ that of monolithic GaAs, which can be explained as in the case of the filling fraction dependence in the range of high filling fraction values. In the atomic structure of AlGaAs a direct bandgap only occurs at low Al concentrations. For the effective mass approximation to be applicable we have confined the range of this parameter to $d < 0.35$. In the range $0.14 < d < 0.35$ the ultimate efficiency increases almost linearly from $0.405$ to $0.425$. This growth is due to the deepening of potential wells with increasing concentration of Al in the matrix. Deeper wells imply a greater distance between the $IB$ and the $VB$ formed by strongly overlapping hole minibands, which tend to lie above the matrix effective potential level. The efficiency increases with growing $E_{I}$ (cf. Fig.~\ref{fig:f2}), due to an increased rate of transitions via $IB$. Carrier excitation in these cascade transitions is mainly hindered by a low flux of absorbed photons able to induce transitions between the $IB$ and the $CB$.

\begin{figure}
\centering
\includegraphics[width=3in]{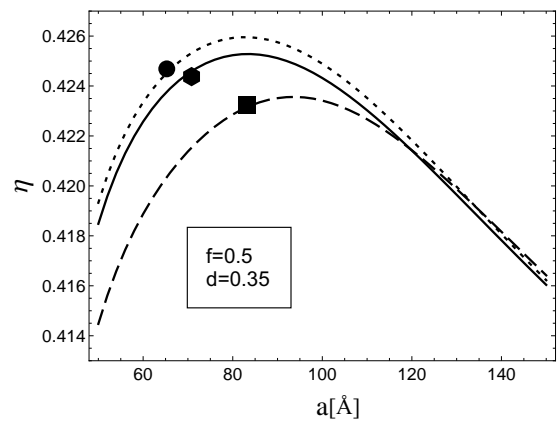}
\caption{\label{fig:f7}Ultimate efficiency $\eta$ versus superlattice constant $a$, as calculated for a filling fraction $f = 0.5$ and a matrix Al concentration $d = 0.35$. Plots obtained for superlattices with rods of square, hexagonal and circular cross section are labeled with respective geometric figures.}
\end{figure}

Plotted in Fig.~\ref{fig:f7}, the dependence on the superlattice constant $a$ has been obtained for the filling fraction fixed at $f = 0.5$ and matrix Al concentration $d = 0.35$. The efficiency is easily seen not to vary widely in the considered lattice constant range, $50 \mathring{A} < a < 150 \mathring{A}$; the maximum difference is $1.2\%$, approximately.

Increasing the superlattice constant $a$ at fixed filling fraction implies widening of both wells (associated with rods) and barriers (associated with the matrix). For small values of $a$ the $IB$ is weakly bound by narrow potential wells, in spite of the strong interactions between them. As a result, the $IB$ lies beyond the well and is relatively wide. The distance $E_{I}$ between the bottoms of the $IB$ and the $CB$ is relatively large; the decrease in efficiency is mainly due to the widening of the gap $E_{G}$ between the $VB$ and the $CB$. For large superlattice constant values the wells are wide and well isolated; in this situation both the miniband spacing and the miniband width are small; a number of non-overlapping minibands can occur in potential wells. Our model assumes a single intermediate band. Omitting higher minibands that do not overlap can lead to an underestimation of the efficiency. However, we believe that taking into account more minibands would not reverse the downward tendency of the efficiency with increasing superlattice constant.

Note in the case of rods of circular and hexagonal cross section the efficiency reaches maximum at a superlattice constant of about $80 \mathring{A}$, while with square cross-section rods the optimum superlattice constant value is $95 \mathring{A}$.

\section{Conclusions}
We have estimated the ultimate efficiency of solar cells based on a two-dimensional superlattice of GaAs rods disposed in sites of a square lattice and embedded in an AlGaAs matrix. The efficiency of such superlattice-based solar-energy converters is found to exceed by a few percent that of monolithic GaAs in a wide range of parameters. The key role in this gain in efficiency is played by the lowest conduction miniband, which, detached from the block of overlapping conduction minibands, acts as an intermediate band that opens an extra channel for carrier transitions between the valence band and the conduction band. The position of this intermediate miniband is determined by the distance $E_{I}$ between its bottom and the bottom of the block of overlapping conduction minibands. Another parameter of vital importance for the efficiency of solar-energy conversion is the distance between the top of the highest valence miniband and the bottom of the block of overlapping conduction minibands. This distance, or bandgap $E_{G}$, determines the energy of utilized carriers.

The ultimate efficiency of a solar cell with intermediate band is found to reach a maximum value of $68\%$ for $E_{G} = 1.95$ eV and $E_{I}/E_{G} = 0.37$. However, this maximum efficiency is unavailable in the material considered in this study, mainly due to the limited distance $E_{I}$ between the intermediate band and the conduction band from which it is detached.

The filling fraction $f$ proves to be the most important factor affecting the efficiency of the studied systems. For $f = 0.45$ the ultimate efficiency exceeds that of monolithic GaAs by $3\%$, approximately. Of lesser impact is the superlattice constant, here allowed to vary in the range from $50\mathring{A}$ to $150\mathring{A}$ (at fixed $f$). Maximum efficiency is attained at a superlattice constant of about $90\mathring{A}$. Interestingly, the geometry of rods is of negligible effect on the efficiency (as long as $f$ is kept constant). The efficiency proves slightly higher in superlattices with rods of more regular shape.

\appendix
\section{}
Defined as the area ratio of the rod cross section to the superlattice cell, the filling fraction ($f_{t}$, $f_{s}$, $f_{h}$ and $f_{c}$, respectively) of superlattices with rods of triangular, square, hexagonal or circular cross section is expressed as follows:
\begin{eqnarray}
f_{t}=\tfrac{\sqrt{3}}{4}\left(\tfrac{l}{a}\right)^{2},&f_{s}=\left(\tfrac{b}{a}\right)^{2},\\
f_{h}=\tfrac{3\sqrt{3}}{2}\left(\tfrac{h}{a}\right)^{2},&f_{t}=\pi\left(\tfrac{r}{a}\right)^{2},
\end{eqnarray}
where $l$, $b$ and $h$ denote the side length of the triangular, square and hexagonal cross section, respectively, and $r$ is the cross-section radius of the cylindrical rods.

Fourier coefficients $F^{\bf G}$ (for ${\bf G}=2\pi /2 (m,n)$) of a periodic function $f({\bf r})$ (taking on the respective constant values $f_{A}$ and $f_{B}$ in the rods and the matrix) in a two-dimensional square lattice of rods of cross section in the form of an equilateral triangle ($F^{\bf G}_{t}$), a square ($F^{\bf G}_{s}$), a regular hexagon ($F^{\bf G}_{h}$) or a circle ($F^{\bf G}_{c}$) are given by the following formulae:
\begin{widetext}
\begin{eqnarray}
F^{\bf G}_{t}&=&\left\{
\begin{array}{l}
f_{B} + (f_{A} - f_{B}) \frac{\sqrt{3} l^{2}}{4 a^{2}}\;\; {\rm for} \;\; m=n=0,\\[1ex]
(f_{A} - f_{B})\frac{1}{6\pi^{2}a n^{2} }e^{-i2\pi n l/\sqrt{3}a}
\left[e^{i\pi n l\sqrt{3}/a}\left(\sqrt{3}a - i 3\pi n l  \right)-\sqrt{3}a\right]\\
 \;\;{\rm for} \; m=0,n\neq 0,\\[1ex]
(f_{A} - f_{B})\frac{1}{2\pi^{2}( m^{3}-3m n^{2})}
 e^{-i2\pi n l/\sqrt{3}a}\left[\sqrt{3}m+e^{i\pi n l\sqrt{3}/a}\left(
-\sqrt{3}m\cos(m\pi l/a)\right. \right. \nonumber\\ \left. \left. +3in\sin(m\pi l/a)
\right)\right]\\
 \;\;{\rm for} \; m\neq 0,n\neq 0,
\end{array}
\right.
\nonumber\\[1.5ex]
%\end{eqnarray}
%\begin{eqnarray}
F^{\bf G}_{s}&=&\left\{
\begin{array}{ll}
f_{B} + (f_{A} - f_{B}) \frac{b^{2}}{a^{2}}&
 {\rm for} \;\; m=n=0,\\[1ex]
(f_{A} - f_{B})\frac{b}{\pi a m}
\sin(m\pi b/a)
&{\rm for}\; m\neq
 0,n=0,\\[1ex]
(f_{A} - f_{B})\frac{b}{\pi a n}
\sin(n\pi b/a)
&
 {\rm for}\; m=
 0,n\neq 0,\\[1ex]
(f_{A} - f_{B})\frac{1}{\pi^{2}m n}
 \left[\sin(m\pi b/a)\sin(n\pi b/a)
\right]&
 {\rm for} \; m\neq 0,n\neq 0,
\end{array}
\right.\nonumber\\[1.5ex]
%\end{eqnarray}
%\begin{eqnarray}
F^{\bf G}_{h}&=&\left\{
\begin{array}{l}
f_{B} + (f_{A} - f_{B}) \frac{3\sqrt{3} h^{2}}{2 a^{2}}\;\; {\rm for} \;\; m=n=0,\\[1ex]
(f_{A} - f_{B})\frac{1}{a\sqrt{3}m^{2} \pi^{2}}
\left[a-a\cos(\sqrt{3}m\pi h/a)+h\sqrt{3}m\pi\sin(\sqrt{3}m\pi h/a)\right]\\
 \;\;{\rm for} \; m\neq 0,n= 0,\\[1ex]
(f_{A} - f_{B})\frac{1}{\pi^{2}( m^{2}n- n^{3})}
\left[\sqrt{3}n\left(\cos(2n\pi h/a)-\cos(\sqrt{3}m\pi h/a)\cos(n\pi h/a)\right)\right.\nonumber\\ \left.+3m\sin(\sqrt{3}m\pi h/a)\sin(n\pi h/a)\right]\\
 \;\;{\rm for} \; m\neq 0,n\neq 0,
\end{array}
\right.\nonumber\\[1.5ex]
%\end{eqnarray}
%\begin{eqnarray}
F^{\bf G}_{c}&=&\left\{
\begin{array}{ll}
f_{B} + (f_{A} - f_{B}) \pi\frac{r^{2}}{a^{2}}&
 {\rm for} \;\; m=n=0,\\[1ex]
(f_{A} - f_{B})\frac{r}{a\sqrt{m^{2} + n^{2}}}\;
{\rm J_{1}}\!\!\left(2 \pi \sqrt{m^{2} + n^{2}}\; r/a\right)
&
 {\rm for}\; m
 \neq,n\neq 0,
\end{array}
\right.
\end{eqnarray}
\end{widetext}
where ${\rm J_{1}}\!(r)$ is a Bessel function of the first kind.

% If you have acknowledgments, this puts in the proper section head.
\begin{acknowledgements}
This study was supported by Polish Ministry of Science and Higher Education, grant No.~N~N507~3318~33.
\end{acknowledgements}
%\vspace{1in}
% Create the reference section using BibTeX:
%\bibliography{basename of .bib file}

\end{document}